\documentstyle[12pt,epsf]{article}
\newcommand{\la}[1]{\label{#1}}

\newcommand{\be}{\begin{equation}}
\newcommand{\ee}{\end{equation}}
\newcommand{\ba}{\begin{eqnarray}}
\newcommand{\ea}{\end{eqnarray}}
\newcommand{\rmi}[1]{{\mbox{\scriptsize #1}}}
 
\newcommand{\nr}[1]{(\ref{#1})}

\newcommand{\roots}{\sqrt{s}}
\newcommand{\fm}{{\rm fm}}
\newcommand{\gev}{{\rm GeV}}
\newcommand{\psat}{p_{\rmi{sat}}}

\def\lsim{\raise0.3ex\hbox{$<$\kern-0.75em\raise-1.1ex\hbox{$\sim$}}}
\def\gsim{\raise0.3ex\hbox{$>$\kern-0.75em\raise-1.1ex\hbox{$\sim$}}}

\begin{document}
 
\begin{titlepage}
\begin{flushright}
21 September 1999\\
JYFL-8/99\\
hep-ph/9909456\\
\end{flushright}
\begin{centering}
\vfill

{\bf 

          SCALING OF TRANSVERSE ENERGIES AND MULTIPLICITIES WITH
              ATOMIC NUMBER AND ENERGY IN ULTRARELATIVISTIC
                     NUCLEAR COLLISIONS 
}

\vspace{0.5cm}
 K.J. Eskola$^{\rm a}$\footnote{kari.eskola@phys.jyu.fi},
 K. Kajantie$^{\rm b}$\footnote{keijo.kajantie@helsinki.fi},
P.V. Ruuskanen$^{\rm a}$\footnote{vesa.ruuskanen@phys.jyu.fi} and
K. Tuominen$^{\rm a}$\footnote{kimmo.tuominen@phys.jyu.fi}

\vspace{1cm}
{\em $^{\rm a}$ Department of Physics, P.O.Box 35, FIN-40351
Jyv\"askyl\"a, Finland\\}
\vspace{0.3cm}
{\em $^{\rm b}$ Department of Physics,
P.O.Box 9, FIN-00014 University of Helsinki, Finland\\}

\vspace{1cm}
{\bf Abstract}

\end{centering}

\vspace{0.3cm}\noindent
We compute how the initial energy 
density and produced gluon, quark and
antiquark numbers scale with atomic number and beam energy in
ultrarelativistic heavy ion collisions. The computation is based
on the argument that the effect of all momentum scales can be
estimated by performing the computation at one transverse
momentum scale, the saturation momentum. 
The initial numbers are converted to
final ones by assuming kinetic thermalisation and adiabatic
expansion. 
The main emphasis of the study is at LHC and RHIC energies but 
it is observed that even at SPS energies this approach
leads to results which are not unreasonable: 
what is usually described as
a completely soft nonperturbative process can also be described in
terms of gluons and quarks.
The key element is the use of the saturation scale.

\vfill 

\end{titlepage}

\section{Introduction} 
In ultrarelativistic heavy ion collisions, the number of produced
gluons and quarks with $p_T$ larger than some cut-off
$p_0$, $N_{AA}(p_0,\roots)$,
increases \cite{bm}-\cite{ekl}  when $p_0$ decreases, when the
size of the nucleus increases ($\sim d\sigma/\sigma_\rmi{inel}\sim
A^2/R_A^2\sim A^{4/3}$) and when $\roots$ increases (since the small-$x$
enhancement \cite{HERA} of the distribution 
functions becomes more effective).
Shadowing will decrease the number \cite{shadowing}, but NLO
corrections \cite{et} will increase it. At sufficiently large cut-off
$p_0\gg \Lambda_{\rm QCD}$, the system of produced gluons and quarks 
is dilute and usual perturbation theory can be applied. However, at 
some transverse momentum scale $p_0=\psat$
the gluon and quark phase space density saturates \cite{glr}-\cite{mq}
and one does not expect further increase. In this case one may
conjecture that the amount of $E_T$ produced at $\psat$ gives a good
estimate of the total $E_T$ produced in an average nucleus-nucleus
collision: gluons with $p_T\gg \psat$ carry lots of $E_T$ but
are very rare, whereas gluons with $p_T\ll\psat$ are numerous
but carry little $E_T$. Doing the computation at $\psat$ gives an
estimate of the effect from all scales, both above and below $\psat$.

In this note, we shall work out the quantitative consequences of the
above conjecture for the amount of 
$E_T$ initially produced in an $A+A$ collision at some large $\roots$.
The key element is the definition and computation of the saturation
scale. For this, one first computes the
number $N_{AA}(p_0,\roots)$ of produced quanta including
perturbatively all gluons and quarks with $p_T>p_0$.
A saturation criterion is formulated and the solution of this
gives $\psat$. Evaluating $N_{AA}(\psat,\roots)$ and 
$E_T^{AA}(\psat,\roots)$ then gives
the initial values.
All quantities can well be fitted to a scaling law of the
type ($\roots$ is in units of GeV in the formulae)
\be
CA^a(\roots)^b,
\la{scaling}
\ee
possibly with an additional term constant in $s$.

This computation at $\psat$ only gives 
initial values at the proper time
$\tau_i=1/\psat$ and not yet experimentally measurable quantities.
To get these, one has to trace the system through the entire set of
expansion, hadronisation and decoupling stages. We shall also discuss
this assuming kinetic thermalisation of the gluonic component and
adiabatic expansion. 

One is accustomed to considering the average events at SPS energies
($\roots\lsim$ 20 GeV) as essentially soft, stringlike. It is somewhat
surprising that this entirely hard, perturbative, treatment will be
seen to give numbers which are not unreasonable
even at these relatively
low energies.  This phenomenon has also been pointed out in
\cite{srivastava1,srivastava2} in the framework of a parton cascade
model.  From the point of view of our study the crucial factor is the
use of the saturation scale as the scale of dominant processes.

There are clearly many refinements (or objections) one can suggest to
this approach. The most obvious is that one should try to spread out
the relevant momentum scales around $\psat$ or, in time, around
$\tau_i=1/\psat$. Further, there certainly will be some entropy
production during the expansion stage -- that due to chemical equilibration
has been estimated in \cite{er}. We will discuss these briefly,
but the key issue anyway is the use of $\psat$ as the dominant scale.

If $\psat$ is very large, well into the perturbative regime, there are
transverse scales $p_T\ll\psat$ which are perturbative but for which
transverse phase space occupation numbers are $\gg1$. A classical field
description is then appropriate \cite{elm}-\cite{kv2}, in analogy with
the computation of the rate of baryon number violating reactions in
electroweak theory \cite{moore}.

\section{Initial values}
The computation is technically a standard pQCD one 
and uses the formulas in Section 
2 of \cite{ek}. The parton distribution functions are from \cite{distr} 
and nuclear effects to them are implemented using the EKS98-parametrization 
\cite{shadowing}. The corresponding
NLO corrections  to $E_T$ within the
appropriate acceptance region ($|y| < 0.5$, all azimuthal angles
around the beam direction, and $p_T>p_0$) 
have been computed in \cite{et}, see also \cite{lo}.
This is an infrared safe computation. 
For, say, $p_0=2$ GeV 
and choosing all the scales as $p_T$, one finds  
$K=1.7$ at $\roots=5500\,\gev$, $K=2.3$ 
at $\roots=200\,\gev$ and  $K=1.9$
at $\roots=20\,\gev$. In principle, the $K$-factor should be computed
at the saturation scale, but in view of this small range of variation
we shall take $K=2$. 
Anyway the NLO corrections to $N$ cannot be
computed in an IR safe manner without introducing some additional 
regulator.

The calculation now proceeds as follows. First compute the average 
number $N_k(p_0)$, $k=g,q,\bar q$, of quanta produced in an average 
central $A+A$ collision within $|y|<0.5$ 
(for brevity, the fact that various quantities may depend on $A$, 
centrality,
$\roots$ or $p_0$ is not explicitly marked in what follows). 
The saturation limit is then obtained from the equation
\be
N(p_0)=\sum_{k=g,q,\bar q}N_k(p_0)=p_0^2R_A^2,
\la{psat}
\ee
where $R_A=1.12\,A^{1/3}$ fm. This expresses the fact that at
saturation $N(\psat)$ quanta each with transverse area $\pi/\psat^2$ fill
the whole nuclear transverse area $\pi R_A^2$. No numerical or group
theory factors nor powers of $g_s$ are included; these are anyway
{\cal O}(1) unless one discusses a parametric weak coupling limit
$g_s\to0$. All parton flavours are included,
though at $p_0=\psat$ gluons clearly dominate even at lowest energies.
The way Eq.~\nr{psat} determines $\psat$ for $A=208$
and $\roots=17,200,5500$ GeV is shown in Fig.~\ref{sqcd}. The numerical
values computed for various $A$ and $\roots$ form a family of curves
shown in Fig.~\ref{nchfig} (normalised to give final charged
multiplicity). They can be
well fitted by (the points with $\roots>200$ GeV are used in determining
the parameters, though the fit agrees well also with the points
computed at lower $\roots$)
\ba
N_i=N(p_0=\psat)&=&1.383\,A^{0.922}(\roots)^{0.383},\la{nscaling}\\ 
\psat&=&0.208\,\gev\, A^{0.128}(\roots)^{0.191},\la{psatscaling}\\ 
\psat N_i&=&0.288\,\gev\,A^{1.050}(\roots)^{0.574}.\la{pN}
\ea

With given assumptions ($K=2$, certain set of parton distributions,
certain method of implementing shadowing, certain scale choices, etc) 
we require the accuracy of the fit to be within 1\% at the fitted region and 
quote the parameters with corresponding accuracy. Clearly there is a 
larger error related to the underlying theoretical assumptions (e.g. the 
$K$-factor) and to the whole starting point: the use of $\psat$ as the 
dominant scale.

\begin{figure}[htb]
\vspace{6cm}
\hspace{0.5cm}
\epsfysize=12cm
\centerline{\epsffile{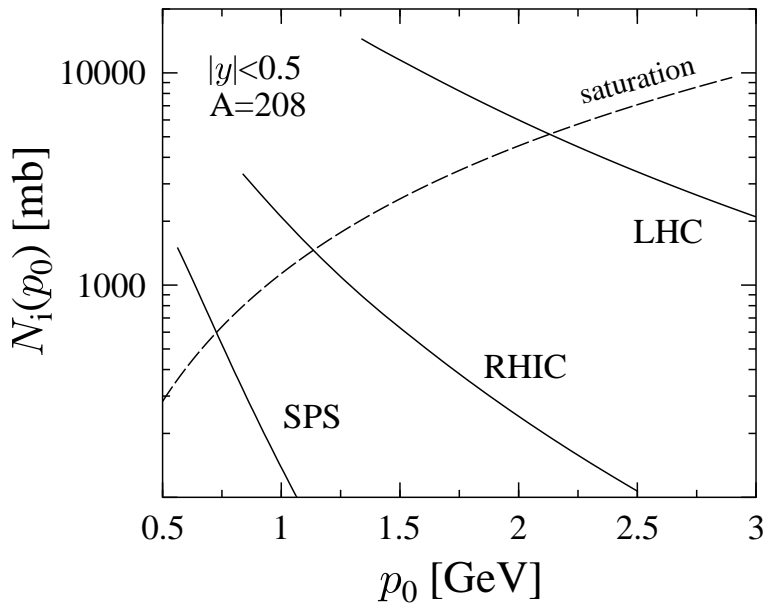}}
\vspace{-8.9cm}
\caption[a]{\protect \small The average number of initially 
produced QCD-quanta 
with $p_T\ge p_0$ and $|y|<0.5$ as a function of the lower limit 
$p_0$ for central Pb-Pb collisions at $\sqrt{s}= 5500$
(LHC), 200 (RHIC) and 17 GeV (SPS).
The saturation scale $p_{\rm sat}$ for $A=208$ is given by the points of 
intersection of the dashed curve ``saturation'' 
($p_0^2R_A^2$) with the curves $N(p_0)$.
}
\la{sqcd}
\end{figure}

Secondly, with the $\psat$ so obtained, the initial values $E_{Ti,k}$
and their sum $E_{Ti}$
can be computed. The results for the sum are plotted in Fig.~\ref{etini} 
and they behave as
\ba
E_{Ti}&=&0.386\,\gev\,A^{1.043}(\roots)^{0.595} \\
&=&1.34\,A^{-0.007}(\roots)^{0.021}\psat N_i.\la{etiscaling}
\ea

By comparing \nr{pN} and \nr{etiscaling}
one observes that approximately $E_{Ti}\approx {\rm const}\,\psat N$.
Actually the dependence on $\roots$ and, in
particular, on $A$ is so weak that one may question whether the 
deviation is a genuine physical effect.
To interpret the result, assume first that precisely
\be
E_{Ti}=C_1\psat N_i, \la{etpsat}
\ee
with $C_1$ = constant.
Then, using the initial volume estimate $V_i=\pi R_A^2/\psat$ and the 
saturation relation $N_i=\psat^2R_A^2$ one obtains
\ba
\epsilon_i&=&{E_{Ti}\over V_i}={C_1\over\pi}\psat^4,\\ 
n_i&=&{N_i\over V_i}={1\over\pi}\psat^3,\\ 
{\epsilon_i\over n_i}&=&C_1\psat.\la{epsovern}
\ea
Since the gluons are by far the dominant component,
even at the SPS, let us convert $\epsilon_i$ 
to a temperature including only them, i.e., writing $\epsilon=16\pi^2/30
\cdot T^4$. Then
\be
T_i=\left({30C_1\over 16\pi^3}\right)^{1/4}\psat.
\la{tpsat}
\ee
Comparing \nr{epsovern} and \nr{tpsat},
the boson gas thermal relation $\epsilon/n=2.70 T$ (=3.15$T$ for fermions)
is seen to hold if
$C_1^3=2.7^430/(16\pi^3)$, $C_1=1.48$; very close to what 
the numerical calculation gives for it (see Eq.~\nr{etiscaling}).

\begin{figure}[htb]
\vspace{4cm}
\hspace{0cm}
\centerline{\epsfysize=9.5cm\epsffile{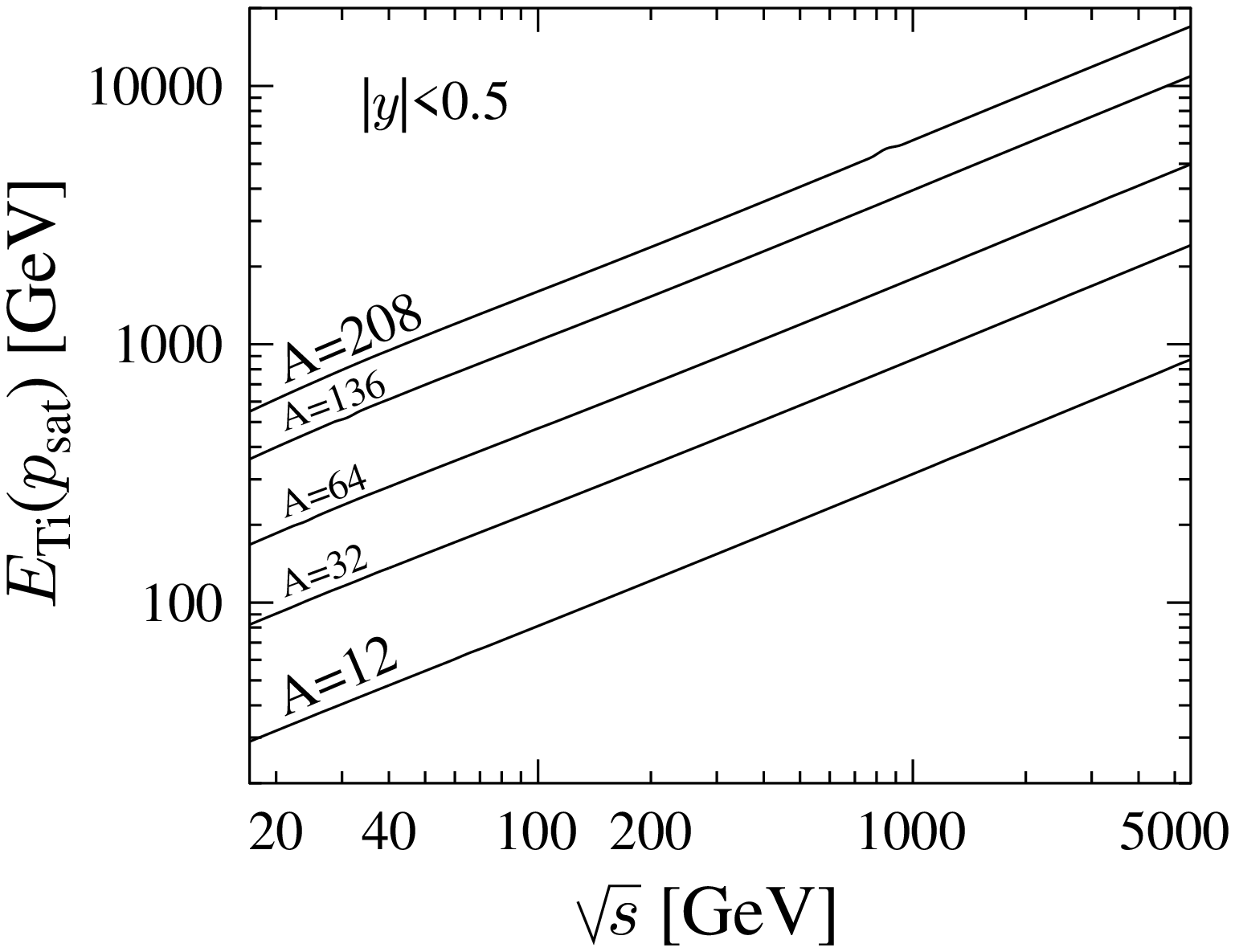} 
\hspace{-1cm}\epsfysize=9.5cm\epsffile{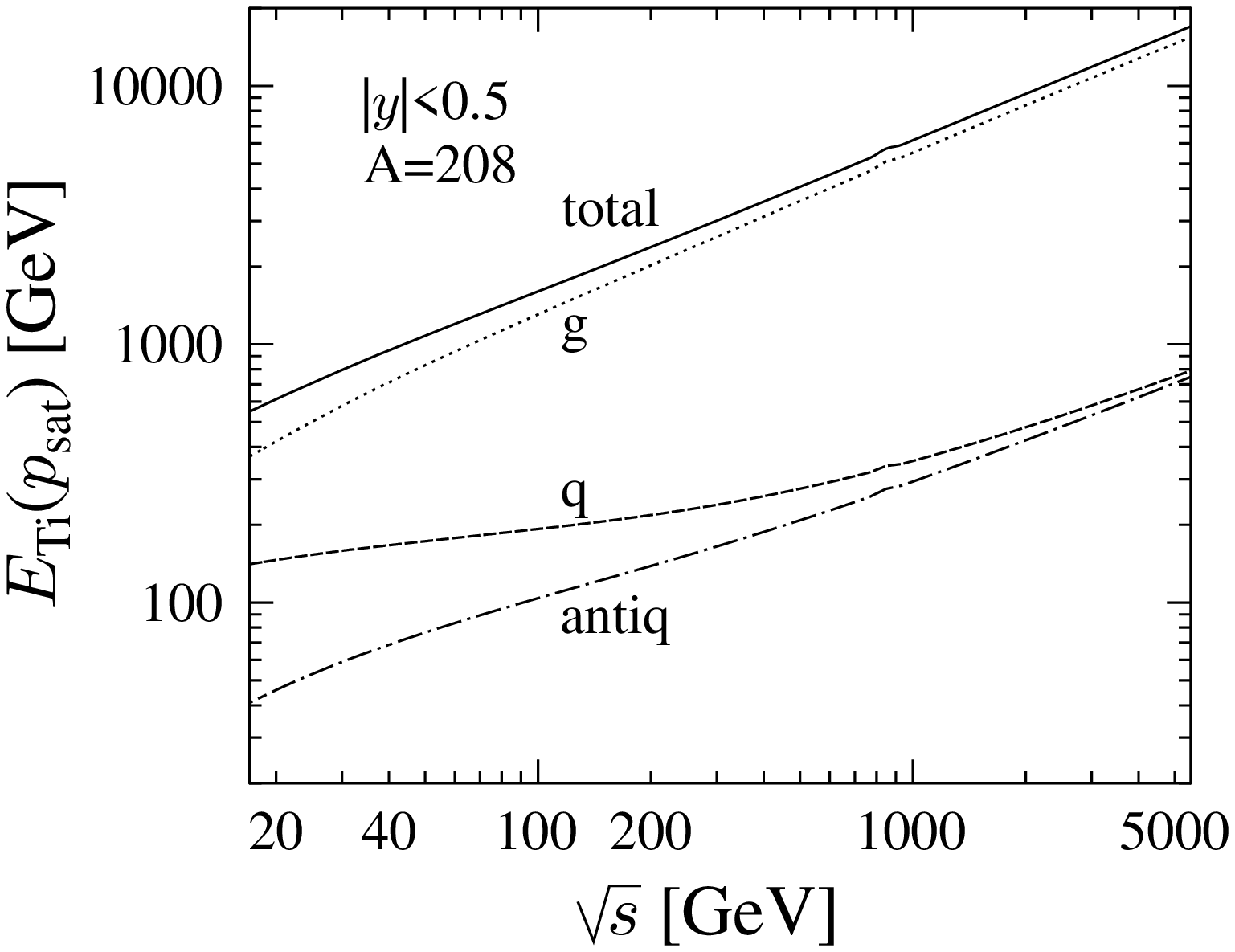} }
\vspace{-7cm}
\caption[a]{\small (a) The initial $E_T$ in $|y|<0.5$ in a central 
$A+A$ collision with $A=12,32,64,136,208$ as a function of 
$\roots$. (b) Decomposition of the initial $E_T$ into
gluon, quark and antiquark components for $A=208$.
}
\la{etini}
\end{figure}

Including the small deviations from $E_{Ti}=C_1\psat N_i$ one finds,
similarly:
\ba
\epsilon_i&=&{E_{Ti}\over V_i}=0.103\,\gev\fm^{-3}A^{0.504}(\roots)^{0.786}
,\\ \la{epsiscaling}
n_i&=&{N_i\over V_i}=0.370\,\fm^{-3}A^{0.383}(\roots)^{0.574},\\
\la{niscaling}
T_i&=&0.111\,\gev\, A^{0.126}(\roots)^{0.197}=
0.534A^{-0.002}(\roots)^{0.006}\psat,
\la{Tiscaling}\\ 
{\epsilon_i\over n_i}&=&2.52\,A^{-0.005}(\roots)^{0.015}T_i.
\la{epsnratio}
\ea
One observes that the ideal boson gas thermalisation relation
$\epsilon/n=2.70T$
holds almost independent of $A$ and $\roots$: the system satisfies this
criterion of thermalisation right at production. Note that this applies
to some degree even at the SPS. 
The scaling parameter values together with some relevant numerical
values are summarised in Table~\ref{parameters}.

\begin{table}
\center
\begin{tabular}{|c|c|c|c|c|c|c|}
\hline
Quantity  &  $C$ & $a$  & $b$   & SPS   & RHIC & LHC   \\
\hline
$N_i$        & 1.383    & 0.922  & 0.383 & 598  & 1440 & 5140     \\
$\psat/\gev$   & 0.208   & 0.128  & 0.191 & 0.73  & 1.13 & 2.13   \\
$E_{Ti}/\gev$   & 0.386   & 1.043  & 0.595 & 600  & 2360 & 17000   \\
$\epsilon_i/(\gev/\fm^3)$ & 0.103  & 0.504  & 0.786 & 16.1 & 98.2 &1330 \\
$n_i\cdot\fm^3$   & 0.370   & 0.383  & 0.574 & 16.0  & 59.8 & 401   \\
$T_i/\gev$   & 0.111   & 0.126 & 0.197 & 0.39  & 0.62 & 1.19   \\
$E_{Tc}$  & 3.477$T_c$  & 0.917  & 0.398 & 276  & 692 & 2600   \\
\hline 
\end{tabular}
\caption[1]{\protect\small Summary of the values of the scaling 
parameters $C,a,b$ in
quantity = $C\,A^a(\roots)^b$, together with numerical values for $A=208$ at
SPS, RHIC and LHC ($\roots=20,200,5500$ GeV). The numbers for $E_{Tc}$ 
(Eq.~\nr{etc}) are for $T_c=0.18$ GeV.}
\la{parameters}
\end{table}

It is also illustrative to consider the results for $g,q,\bar q$
separately.  The $A$ and $\roots$ dependences of the different
components are almost the same but to account for the growing
importance of the valence quark component towards smaller energies,
one also has to include a constant term in the fit. For simplicity, we
first make a four paramenter fit $A^a[C(\roots)^b+D]$ to the total
initial multiplicity $N_i$ for $\roots>40$ GeV.  With the powers $a$
and $b$ so fixed, we find $C$ and $D$ separately for each
component. The result is
\ba
N_g&=&A^{0.922}[1.065(\roots)^{0.404}-0.028],\\\nonumber
N_q&=&A^{0.922}[0.021(\roots)^{0.404}+0.778],\\\nonumber
N_{\bar q}&=&A^{0.922}[0.037(\roots)^{0.404}+0.287],
\ea
and for the initial transverse energy
\ba
E_{Tg}&=&A^{1.044}[0.341(\roots)^{0.599}-0.398],\\\nonumber
E_{Tq}&=&A^{1.044}[0.0137(\roots)^{0.599}+0.494],\\\nonumber
E_{T\bar q}&=&A^{1.044}[0.0153(\roots)^{0.599}+0.153].
\ea
Note that the simple parametrisations above are only meant to reproduce 
the behaviour of the different components, not the
small difference  $N_q-N_{\bar q}$. They cannot be used
to compute, e.g., the net quark number at the LHC, 
since $(N_q-N_{\bar q})/N_g <$ the  accuracy of the fits.

\section{Expansion stage}
We have now the initial values at $\tau_i=1/\psat$. To compare with
experiment one has to follow the evolution of the system through expansion
in the QCD plasma phase, phase transition, expansion in the hadronic
phase and decoupling. Now that one has computed from perturbative
QCD that the initial state very nearly satisfies the kinetic
thermalisation condition $\epsilon/n=2.7\,T$, there is some justification in
assuming that also further expansion is locally thermal, i.e.,
entropy conserving. There is a large number of dissipative effects
one can think of, but adiabatic boost invariant 
expansion is the baseline here.

\begin{figure}[htb]

\vspace{5cm}
\hspace{0.5cm}
\epsfysize=11cm
\centerline{\epsffile{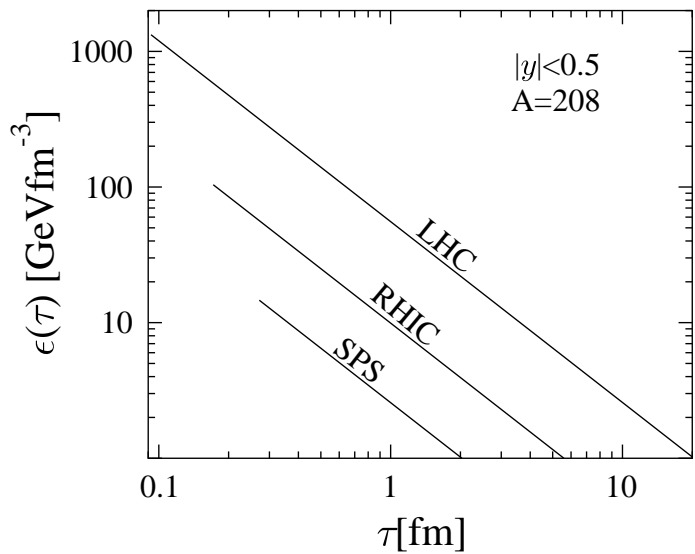}}
\vspace{-8cm}
\caption[a]{\protect \small Proper time dependence of energy density 
during longitudinal expansion for LHC, RHIC and SPS
with the  initial values given in Table~\ref{parameters}.
}
\label{evol}
\end{figure}

Numerical examples of adiabatic expansion are shown in Fig.~\ref{evol}
for the initial one-dimensional stage 
during which $\epsilon(\tau)\tau^{4/3}$ =
constant. We start the expansion at the time $1/\psat$, although at
lower energies this may be smaller than the transit time
$2R_A/\gamma=4R_A/\roots$ of the nuclei. From the scaling formulas it
actually follows that $2R_A/\gamma<1/\psat$ for
\be
\roots > 6.81\,A^{0.570}\,\,\, (=143\,\,{\rm for}\,\,A=208).
\ee
We remind that the whole production process is effectively
described as taking place at the time $1/\psat$.

Later on, at $\tau\sim \sqrt3R_A$, the system will enter a 
regime of longitudinal and transverse expansion,
which has to be studied numerically \cite{kataja}. 
In Fig.~\ref{evol}
we see that at the LHC the transverse expansion effects are likely to set in 
already in the plasma phase.

\begin{figure}[htb]

\vspace{5cm}
\hspace{0.5cm}
\centerline{\epsfysize=11cm\epsffile{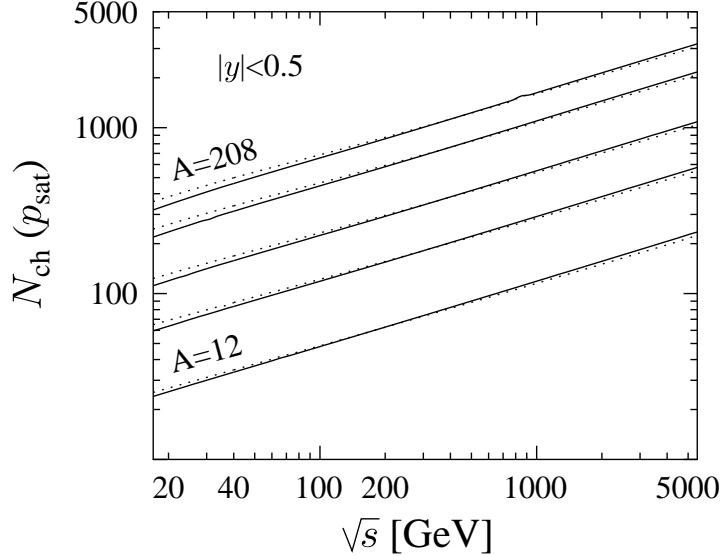}}
\vspace{-8cm}
\caption[a]{\protect \small The number of charged particles 
per unit rapidity for $A=12,32,64,136,208$ as a 
function of $\roots$, computed from Eq.~(\ref{nchlaw}) (solid lines) and 
compared with $2/3*0.9N_i$ (dotted lines). 
}
\label{nchfig}
\end{figure}

For the multiplicity the situation is simple: for an ideal system 
of bosons and fermions the total
entropy is $S=3.60N_B+4.20N_F$ and this is constant per unit rapidity,
as long as the expansion is boost invariant. 
(By definition, all quantities are always for $|y|<0.5$.)
Thus, initially, $S_i\approx3.6N_i$. For the final hadronic
gas $S_i=S_f\gsim4N_f$ so that $N_f=0.9N_i$: the number of
final state hadrons is to up to 10\% corrections equal to 
that of the initially produced partons (gluons)
at the scale $\psat$. The multiplicity prediction 
$N_\rmi{ch}=2/3*0.9N_i$ thus is directly given by Eq.~\nr{nscaling}
and is shown in Fig.~\ref{nchfig}.

Since NLO corrections to $E_T$ can be computed in an IR safe manner,
one might prefer to compute the final entropy from 
the initial $E_T$ through the conversions 
$E_{Ti}\to\epsilon_i\to T_i \to S_i=S_f$.
By making use of the fit (\ref{Tiscaling}), we obtain the following scaling
law for the multiplicity of charged particles per unit rapidity:
\be
N_{ch} = \frac{2}{3}N_f \approx \frac{2}{3}\frac{S_i}{4}= 
\frac{2}{3} 1.16A^{0.92}(\roots)^{0.40}.
\la{nchlaw}
\ee
In view of the observed initial near thermalisation (Eq.~\nr{epsnratio}) 
this necessarily leads to the essentially same scaling in $A$ and $\roots$
as that obtained from $N_i$ in Eq.~(\ref{nscaling}). 
The comparison is shown in Fig.~\ref{nchfig}.

For $E_T$ the situation is more complicated.  
$E_T$ per unit rapidity decreases by the factor $T/T_i$ due to work 
done against expansion \cite{kataja,gpz}. On the other hand, the transverse
expansion at later times will compensate for this effect.

\begin{figure}[htb]

\vspace{5cm}
\hspace{0.5cm}
\epsfysize=11cm
\centerline{\epsfysize=10cm\epsffile{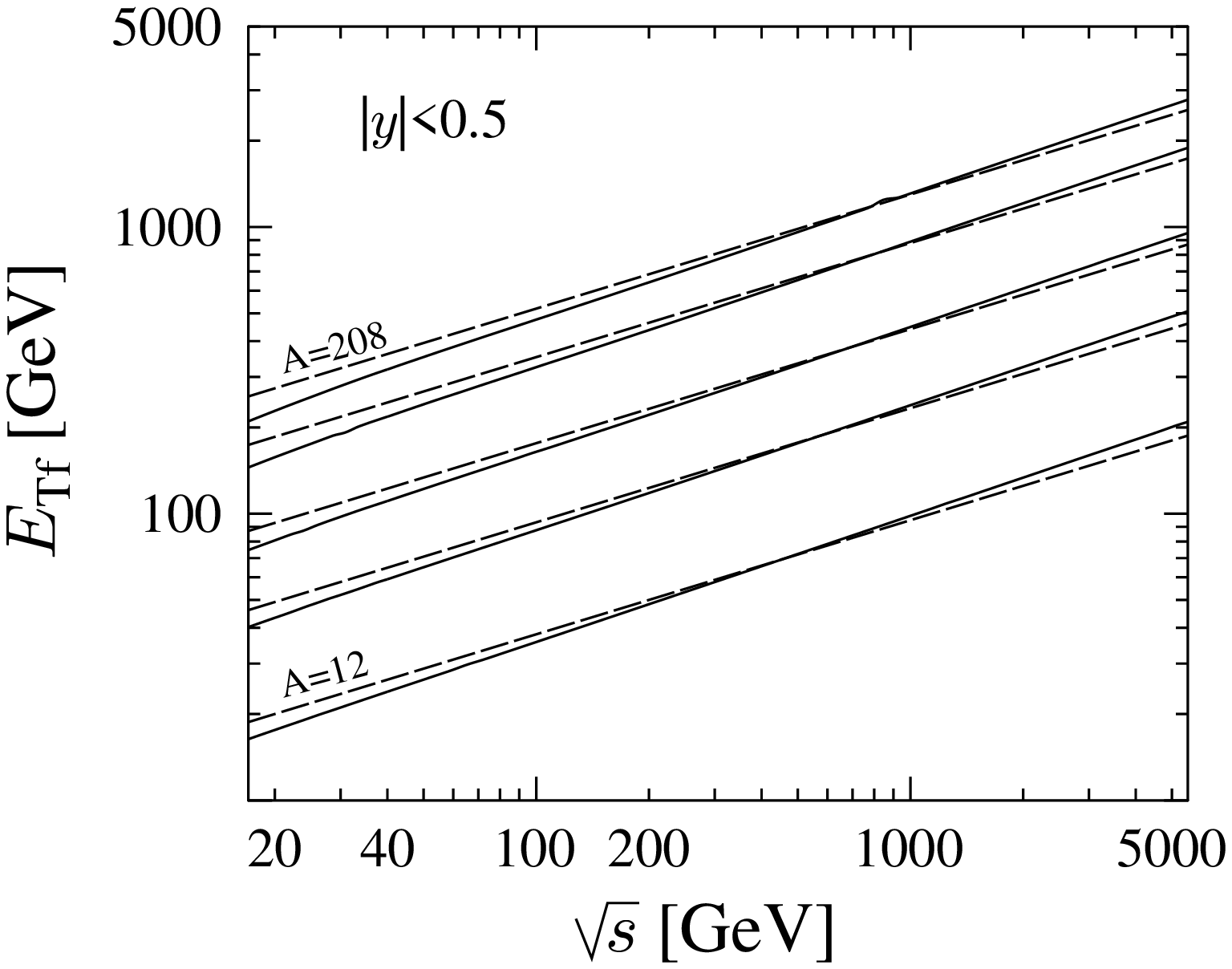} 
\hspace{-1cm}
\epsfysize=10cm\epsffile{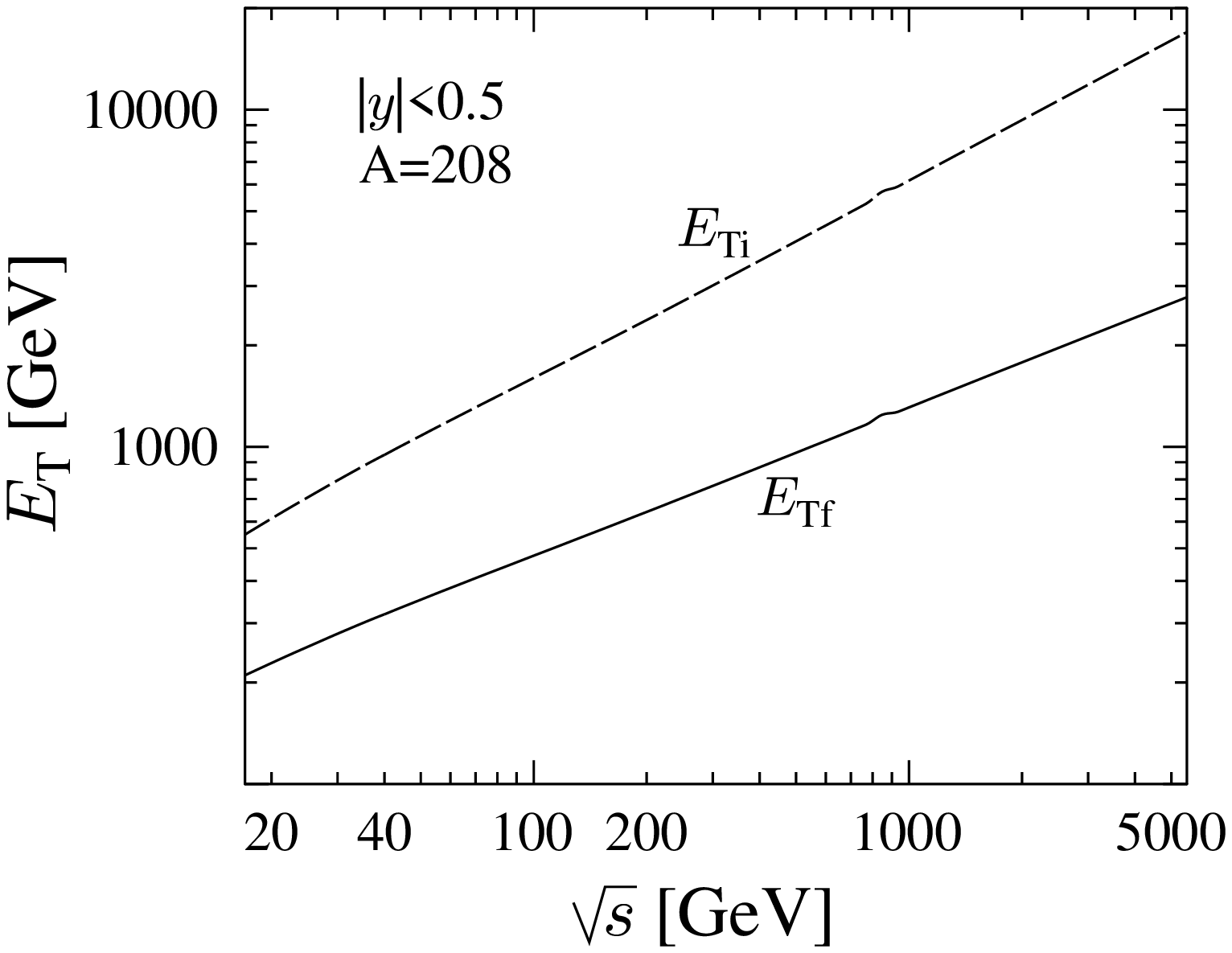} }
\vspace{-7cm}
\caption[a]{\protect\small (a) The average final $E_T$ computed from 
Eq.~\nr{ET_kataja} (solid lines) and from Eq.~\nr{etc} with $T_c=0.18$ GeV 
(dashed lines) for $A=12,32,64,136,208$ as a function of $\roots$. 
(b) Comparison of the initial and the final $E_T$
for $A=208$ as a function of $\sqrt s$.
}
\label{ETfin}
\end{figure}

To estimate the final $E_T$, we extract an approximate scaling law
between the final average transverse momentum $\langle p_T\rangle$ and
the final multiplicity $N_f$ from Fig.~7 of Ref.~\cite{kataja}:
\be
\langle p_T\rangle /{\rm GeV} = 0.39+0.061\ln(N_f/A).
\ee 
By substituting our total multiplicity $N_f$, the final transverse energy 
can then be computed as
\be 
E_{Tf}=N_f \langle p_T\rangle. 
\la{ET_kataja} 
\ee
The transverse energies so obtained are shown in Fig.~\ref{ETfin}.
Using Eq.~\nr{nchlaw} we get the scaling in $A$ and $\roots$:
\be 
E_{Tf} = 0.46 A^{0.92}(\roots)^{0.40}[1-0.012\ln A+0.061\ln\roots].
\la{etfscaling} 
\ee 

On the other hand, a simple estimate of the transverse energy $E_{Tc}$ at 
the end of the plasma phase is obtained by neglecting the transverse expansion 
during the plasma phase. This gives
\ba
E_{Tc}=(T_c/T_i)E_{Ti}&=& 3.477 T_c\, A^{0.917}(\roots)^{0.398}
\la{etc}\\\nonumber
&=&2.51 T_c\,A^{-0.005}(\roots)^{0.015}N_i.
\ea
Using here $T_c=0.18$ GeV, $E_{Tc}$ approximates the 
final $E_T$ computed above very well. The comparison is shown in 
Fig.~\ref{ETfin}a.

Eq.~\nr{etc} expresses a trivial fact: as long as the system is
thermal, the energy per particle is $\sim 2.7T$ - possibly even larger if
also quarks enter chemical equilibrium. Evolution in the hadronic
phase until decoupling will lead to a further reduction by the ratio
$T_\rmi{decoupling}/T_c\lsim1$, but this is compensated for by the
development of flow. Thus the net effect of the hadronic phase can be
expected to be small, and $E_{Tf}\approx E_{Tc}$. 

In Fig.~\ref{ETfin}b we show the difference between the initially 
produced $E_{Ti}$ and the final, measurable, $E_{Tf}$ for $A=208$. At 
the LHC the transverse energy drops by a factor $\sim6$ 
and at RHIC by a factor $\sim 3.5$ due to $pdV$ work
done against expansion. 

\section{Discussion}
\la{discussion}
In the saturated scaling limit ($\sigma\sim\psat^{-2}$)
one expects \cite{bm} the following $A$-scaling rules:
First from Eq.~\nr{psat},
$N\sim A^{4/3}/\psat^2 = (\psat R_A)^2$, one obtains
$\psat\sim A^{1/6}$ and $N\sim A$. Further, $E_{Ti}\sim \psat N \sim
A^{7/6}$. Due to shadowing the observed $A$-exponents 
are somewhat less than these numbers,
for $N_i$ 0.922 instead of 1 and for $E_{Ti}$ 1.043 instead of 1.167.
Anyway it is quite interesting to note these rather small exponents
for this hard process. Note that for a
fixed momentum scale without saturation
the growth with $A$ is much more rapid:  $N_i\sim E_{Ti}\sim A^{4/3}$.

The growth with beam energy is rather rapid, the power of $\roots$ is
$b\approx 0.19$ for $\psat$, 
$b\approx0.38$ for $N_i$ and about the same for the final transverse
energy, depending on the flow effects. A constant valence quark component
can also be identified. The power of $\roots$ depends on the phase
space available and on the small-$x$ increase of the gluon distribution
function.

The decrease of transverse energy in adiabatic expansion
(Fig.~\ref{ETfin}b) is rather large, a factor $\sim 6\,(3.5)$ at the 
LHC (at RHIC). Provided that one can control flow effects numerically,
a measurement of $E_{Tf}$ will give further
constraints on the degree of thermalisation in the system.

An interesting property of saturation calculations appears when
considering them at SPS energies. Usually a heavy ion collision at the
SPS, $\roots\lsim20$~GeV, is regarded as entirely soft, nonperturbative,
dominated by stringlike effects.  Here we have gone to the other
extreme and have treated it in terms of the gluon and quark
degrees of freedom at a very specific scale,
$p_T=\psat$.  This scale is only 0.73 GeV for Pb+Pb collisions 
at the SPS, and perturbation
theory at this scale, although $\alpha_s(\psat)/\pi$ is clearly $<1$,
is not expected to be very accurate (though it works well for, say,
$\tau$ hadronic decays with $m_\tau=1.8$ GeV \cite{neubert}).
However, the numbers one obtains (Table~\ref{parameters}) are not very
far from the measured ones. The charged multiplicity per unit $y$
(actually $\eta$), is experimentally
\cite{na49ch} about 400 for Pb+Pb at $\roots=17$~GeV while 
we obtain $N_\rmi{ch}=330...340$.
For S+S at $\roots=20$~GeV, the experimental value is $N_\rmi{ch}\approx50$ 
\cite{na35ch} while we get $N_\rmi{ch}\approx62...64$, 
again not  dramatically different. 
The average $E_T$ for Pb+Pb is observed to be about 400 GeV \cite{na49et},
somewhat larger than the numbers in Table~\ref{parameters}. We have
not in any way attempted to improve the quantitative agreement with
present data. Maybe some kind of duality is at work here.

It is also notable that even at the SPS the system at $\psat$ is
dominantly gluonic: at $\roots=17$ GeV the gluons take 65\% of $E_{Ti}$
and 73\% of $N_i$. A related fact is that even at the SPS the 
thermalisation condition $\epsilon=2.7Tn$ is initially nearly satisfied.

The RHIC numbers ($A=208$) are $\approx$1300 for the total multiplicity 
$dN/d\eta$ (870 for $N_\rmi{ch}$ from \nr{nchlaw})
and $\approx$660 GeV for $dE_T/d\eta$
and will soon be measured. However, one can compare them with some
historical numbers. In \cite{bm} the values $\psat=0.95$ GeV and
1300 estimated for the multiplicity are very close to the ones 
computed here, but the discussion of $E_T$ included no $p\,dV$ work.

For the LHC, we get the charged particle multiplicity 
$dN_{ch}/d\eta\approx 3300$ and $dE_T/d\eta\approx 2900$ GeV.
These are some 25\% larger than the LHC-numbers in 
Ref. \cite{ek}. 
Now different,  largely compensating, effects are included:
shadowing and a slightly larger value of $p_{\rm sat}$
decrease the numbers while $K=2$ instead of $K=1$ increases them. 

Finally, some remarks are listed:

\begin{itemize}
\item The saturation criterion \nr{psat} has been presented here in a very
simple form. Fundamentally, it is the result of a complicated dynamical
computation containing numerical and group theory factors and powers
of $g_s$ \cite{mq}. The precise value of all these 
will affect the final numbers. Taking the net effect to be = 1 in
\nr{psat} is a geometric estimate based on the uncertainty principle
and there is no way of giving a controlled estimate of the error.

\item The saturation criterion could also be formulated by applying it
to the $p_T$-distribution, calculating it perturbatively for $p_T>\psat$
and assuming that it is, say, constant for $p_T<\psat$. This would lead
to somewhat different numbers, but the error is like that under the
previous point.

\item
To relate the present study to Ref. \cite{emw}, it is perhaps
interesting to compute the screening mass $m_g=gT$ at $T=T_i$. Using
Eq.~\nr{Tiscaling} we get $m_g=g_s(p_\rmi{sat})T_i =
1.6\ln^{-1/2}(p_{\rm sat}/\Lambda_{\rm
QCD})A^{-0.002}(\roots)^{0.006}p_{\rm sat}$. With the
saturation scales in the region $p_{\rm sat}=1...2$ GeV, we get 
$m_g\approx p_{\rm sat}$, since $p_\rmi{sat}\approx 0.5 T_i$.

\item The saturation criterion is formulated for quanta {\em produced} in
the collision. The wave functions of the initial nuclei need not be
saturated for the values of $A$ and $\roots$ considered 
here.

\item The starting point of these calculations of initial production of
QCD quanta in ultrarelativistic central heavy ion collisions 
was that it is
enough to consider one momentum scale, $p_T=\psat$. The magnitude of
the various quantities was then computed using methods appropriate 
for $p_T\ge\psat$.
One can, as well, attempt to use the range
$p_T\le\psat$. This is a region of QCD in which genuinely new
dynamical methods have to be developed. 
Numerical calculations
also there are under way \cite{kv,kv2}. However, whatever the
method applied, it anyway applies only to the initial production
($1/\psat<\tau<1$ fm), leaving the treatment of the entire
further evolution of the system open.

\item There are many interesting and relevant questions concerning the
subsequent flow, after initial production. The key issue is thermalisation,
that of quarks, in particular \cite{er,bdmtw}. 
One could formulate the problem with all degrees of freedom and with
some suppression factors (fugacities \cite{bdmtw}) to account for the
initial deficit of the quarks.  To avoid this complication but to
partly account for the quarks, we used the total
energy density, including quarks, to find the
initial temperature. The entropy increase associated with
chemical equilibration of quarks was estimated to be about
15\% at LHC in \cite{er}.
For SPS one has the problem of the long transit time.

\end{itemize}

\section{Conclusions}
We have in this article computed how the initial values of the energy and
number densities of QCD matter produced in ultrarelativistic 
central heavy ion ($A+A$) collisions depend on $A$ and $\roots$. 
These values were converted to final observed numbers for
$dN_\rmi{ch}/d\eta$ and $dE_T/d\eta$ by assuming adiabatic expansion.
This assumption is justified by the observation that the initial
values very closely satisfy the kinetic thermalisation condition
$\epsilon=2.7nT$ for a dominantly gluonic system at all $A,\roots$.

The computation is technically one in perturbative QCD,
leading order with the NLO corrections estimated by a constant $K$-factor, 
and including nuclear shadowing, but also nonperturbative physics
enters indirectly in that the computations are performed at one
transverse momentum scale, the saturation momentum $\psat$. This is
determined by a geometric transverse saturation criterion and is
supposed to effectively represent all scales, both those $>\psat$
but also those $<\psat$. In time, the initial production is assumed
to effectively take place at $\tau=1/\psat$, instead of a gradual
dumping of energy and entropy from the collision lasting until
$\sim1/\Lambda_\rmi{QCD}$. 

Even at the SPS, where one normally expects soft stringlike physics
to dominate, this entirely hard effective approach gives rather
reasonable numbers. Maybe this gives some credibility to the
predictions for RHIC and LHC, where one expects the approach to
work better. From a purely theoretical point of view one
can think of corrections from many different sources.

{\it Acknowledgements:} We thank L. McLerran and Al Mueller for
discussions, B. M\"uller for extrapolating our curves to a region
where we first dared not go and
the Academy of Finland for financial support.

\end{document}